# Spatiotemporal Arbitrage of Large-Scale Portable Energy Storage for Grid Congestion Relief

Guannan He[1], Da Zhang[2], Xidong Pi[1], Qixin Chen[3], Soummya Kar[1], and Jay Whitacre[1]
1. Carnegie Mellon University; 2. Massachusetts Institute of Technology; 3. Tsinghua University
Email: guannanh@andrew.cmu.edu, zhangda@mit.edu, xpi@andrew.cmu.edu, qxchen@tsinghua.edu.cn, soummyak@andrew.cmu.edu, whitacre@andrew.cmu.edu

*Abstract*—**Energy storage has great potential in grid congestion relief. By making large-scale energy storage portable through trucking, its capability to address grid congestion can be greatly enhanced. This paper explores a business model of large-scale portable energy storage for spatiotemporal arbitrage over nodes with congestion. We propose a spatiotemporal arbitrage model to determine the optimal operation and transportation schedules of portable storage. To validate the business model, we simulate the schedules of a Tesla Semi full of Tesla Powerpack doing arbitrage over two nodes in California with local transmission congestion. The results indicate that the contributions of portable storage to congestion relief are much greater than that of stationary storage, and that trucking storage can bring net profit in energy arbitrage applications.**

## I. Introduction

The increasing deployment of energy storage has driven the storage cost down due to economies of scale [1],[2], which in turn encourages more deployment. There are various storage applications including energy arbitrage [3], reserve [4], frequency regulation [5],[6], renewable integration [7], voltage support [8], etc. Storage can also provide multiple services at the same time [4],[9],[10].

Grid congestion has always been a major concern in power system [10]. Besides transmission capacity expansion, energy storage provides another feasible solution to the congestion problem. The potential use of battery storage to increase transmission capability in thermal-limited transmission lines is investigated in [11]. In [7], a method to coordinate energy storage and wind power plant is proposed to avoid wind curtailment due to transmission congestion and is applied to the grid in northern Chile. In [12], a power flow model for the Pacific Northwest area is built to identify transmission line congestion and analyze the potential contributions of energy storage in certain locations to reducing the congestion. In [13], transmission congestion relief is designed as an ancillary service, and financial compensation is provided to incentivize the contribution of privately-owned storage to congestion relief. The concept financial storage right is proposed in [14] by developing the analogy between energy storage and transmission line. In [10], a multistage model is proposed to coordinate storage in transmission-level congestion relief and distribution-level cost minimization. These studies explore the use of storage in grid congestion relief, although they all consider stationary energy storage only.

The capability of energy storage to relieve congestion is limited by its capacity. When the storage is full or empty after a cycle, it can no longer make any contribution. However, if a storage device is portable and can be transported between nodes with congestion, it can make much greater contributions to congestion relief compared to a stationary storage device at the same size, especially in cases where the congestion is local (the two nodes connected by congested line are close to each other geographically) as the travel time can be very short.

Regarding energy storage transportation for congestion relief, distributed electric vehicle charging control is studied in [15],[16], and vehicle-to-grid application of railway system is studied in [17],[18]. In these studies, there is no active control of the vehicle location for grid congestion relief, and grid congestion relief is just a side benefit. No large-scale storage transportation for grid application has been studied so far.

In this paper, we explore a business model of large-scale portable storage, which consists of truck, energy storage, and power electronics. In this business model, the truck is loaded with storage and inverters and travels between nodes with congestion, which is named as "spatiotemporal arbitrage". We propose a spatiotemporal arbitrage model that determines the optimal operation and transportation schedules of portable storage. In the case study, we apply the spatiotemporal arbitrage model to simulate the schedules of a truck of portable storage using Tesla Semi and Powerpack over two nodes (3 miles away) with local congestion in California. By comparing the life-cycle revenue of portable storage versus stationary storage, we find that trucking storage can not only greatly increase the contribution of storage to congestion relief but also bring net arbitrage profit to storage owner.

The phrase "spatiotemporal arbitrage" is also used in [19] but has a totally different meaning; the spatial decision in [19] is storage siting, which occurs at the long-term planning stage, while the spatial decision in this paper is storage transportation, which occurs in real-time operation stage.

The rest of the paper is structured as follows: the grid congestion in California is discussed in Section II; the concept of portable energy storage is introduced in Section III; the spatiotemporal arbitrage model is introduced in Section IV; Section V presents a case study; Section VI concludes the paper.

## II. LOCAL GRID CONGESTION IN CALIFORNIA

Local transmission congestions have been observed in California recently. There are significant differences in the locational marginal prices (LMP) between some nodes with very short geographical distance, and some occur very frequently. Fig. 1 presents the frequency of price difference between two nodes around San Marcos that are only 3 miles away (5-mile drive) from October 2017 to September 2018. Over the 1-year period, there are over 500 hours when the price difference is greater than $50/MWh for this pair of nodes.

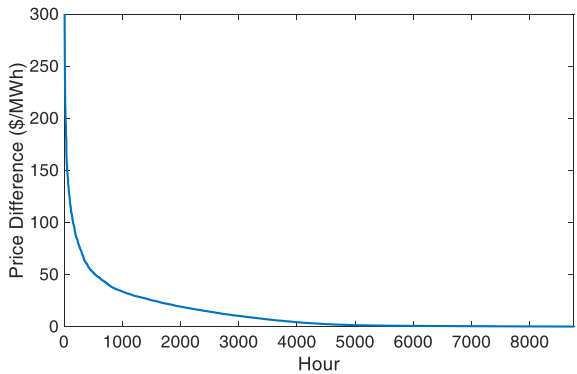

Figure. 1. Frequency of price difference between two nodes around San Marcos (California, US) from October 2017 to September 2018. Node ID: NCMETER_1_N001 and SNTAMRA_1_N005.

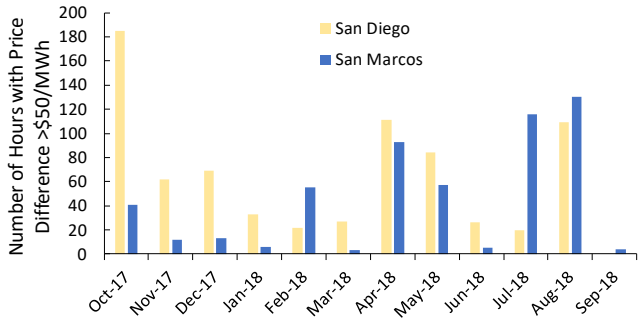

Figure. 2. Monthly distributions of the number of hours when the price difference is greater than $50/MWh for two nodes around San Marcos, CA (Node ID: NCMETER_1_N001 and SNTAMRA_1_N005) and two nodes around San Diego, CA (Node ID: CHCARITA_1_N012 and FRIARS_1_N009).

Is there similar temporal pattern for congestions between different pairs of nodes? No obvious pattern from Fig. 2, which presents the monthly distributions of the number of hours when the price difference is greater than $50/MWh for two pairs of nodes in California, one around San Marcos, and the other around San Diego (15 miles away). There are several months

TABLE I
BASIC PARAMETERS OF LARGE-SCALE PORTABLE ENERGY STORAGE USING TESLA SEMI AND POWERPACK

| | |
|---|---|
| Tesla Semi payload [a] | 35 tonne |
| Energy density of Tesla Powerpack with inverter [b] | 13 kg/kWh |
| Total energy capacity per truck | 2.7 MWh |
| Capital cost of Tesla Semi [a] | $150,000 |
| Energy consumption rate of Tesla Semi [a] | <2 kWh/mile |
| Energy consumption per unit delivered energy for 5-mile roundtrip | 0.7% |
| Labor cost per unit delivered energy for 5-mile roundtrip | $3/MWh |

[a] Source: https://www.tesla.com/semi;
[b] Source: https://www.tesla.com/powerpack;

when both pairs of nodes have comparatively high frequency of significant price difference, for example, April, May, and August in 2018. However, there are also months when one pair has comparatively high frequency while the other does not, for example, November 2017 and February 2018. The correlation coefficient between the two frequency series in Fig. 2 is 0.34, which indicates only weak dependency for the congestions. The absence of uniform congestion pattern may favor the portable storage option as the storage can be shared among different pairs of nodes to make more contributions to the system and earn more profits.

For transmission lines with infrequent congestion, expanding transmission capacity is also less favorable than portable storage, because the investment efficiency is low for the new transmission capacity with low utilization rate. In contrast, when there is no congestion at one location, portable storage can provide other services and serve other locations.

## III. PORTABLE ENERGY STORAGE

To address transmission congestion, a typical solution is to expand transmission line capacity. However, constructing new transmission line is expensive and time-consuming. Instead, there may be another feasible option, which is to use portable large-scale energy storage to transport energy. The basic concept is to load high-energy-density energy storage (lithium-ion batteries for example) and inverters on trucks to mobilize energy storage. The trucks travel between substations with significant price difference, discharging at the high-price node and charging at the low-price node, which we name as spatiotemporal arbitrage.

Considering both environmental benefit and fuel economy, we investigate the case of using Tesla Semi, an electric truck model, to truck Tesla Powerpack. The parameters of Tesla Semi and Powerpack are listed in Table I. Based on the payload of Tesla Semi and the energy density of Tesla Powerpack, we estimate that one truck can accommodate approximately 2.7 MWh batteries with inverters, at a total cost of $2 million. Truck cost accounts for approximately 8% of the total capital cost.

As for variable operational cost, the energy consumption is minimal for short trips between nodes that are geographically close, which accounts for less than 1% of the energy capacity of a truck for 5-mile roundtrip. The labor cost is a considerable cost at $3/MWh, considering a $20/hour wage for truck driver and an average speed at 25 miles/hour.

## IV. SPATIOTEMPORAL ARBITRAGE MODEL

In this section, on top of temporal arbitrage, we develop a spatiotemporal arbitrage model for portable energy storage to maximize arbitrage profit between two nodes subject to operation and transportation constraints.

### A. Objective function

The objective of the spatiotemporal arbitrage dispatch model is to maximize the total market revenue of the portable storage between two nodes ($R_t$) minus the transportation cost ($C_t^{tr}$) and minus the degradation cost ($C_t^d$), as (1). The time horizon of each dispatch decision is typically a day, represented by $\Delta t$, and $t$ is the day index. The decision variables are the discharging and charging schedules of the storage, $P_{n,h}^{dis}$ and $P_{n,h}^{cha}$, and transportation schedules $\gamma_h$. $h$ is the time index, and $\Delta h$ is the dispatch time scale, which typically ranges from 5 minutes to 1 hour.

$$Y_t = \max_{P_{n,h}^{dis}, P_{n,h}^{cha}, \gamma_h} R_t - C_t^{tr} - C_t^d \quad (1)$$

The market revenue of storage is expressed as (2), where $\lambda_{n,h}$ is the forecasted LMP at node $n$ and time $h$; and $n_A$ and $n_B$ denote the two nodes that the storage travels between.

$$R_t = \sum_{h \in [t, t+\Delta t]} \sum_{n \in \{n_A, n_B\}} \left[ \lambda_{n,h}(P_{n,h}^{dis} - P_{n,h}^{cha})\Delta h \right] \quad (2)$$

The main transportation cost is the labor cost, which is assumed to be proportional to the total travel time during day $t$, as (3), where $c_{tr}$ denotes the transportation cost per unit time, and $\gamma_h$ is a 0-1 variable that denotes whether the storage in travel at time $h$.

$$C_t^{tr} = c_{tr} \sum_{h \in [t, t+\Delta t]} \gamma_h \Delta h \quad (3)$$

The degradation cost is an opportunity cost that reflects the loss of future profit opportunity due to current storage usage[9]. The storage degradation can be divided into two categories according to dependent factors [20],[21]: 1) cycling degradation that mainly depends on the amount of energy throughput the storage has processed, as the first expression in (4); and 2) calendar degradation that mainly depends on state of charge (SOC), temperature, and the length of time the storage has experienced [20],[22],[23]. If the average SOC and temperature are assumed to be constant, the calendar degradation during a certain period of time can also be regarded as constant, denoted by $q_t$ in (4). $c_t^d$ is the unit degradation cost, named as the marginal cost of usage, to be determined by the future profitability through life-cycle operation simulations [9]. Typically, the higher the future potential profit, the greater the marginal cost of usage.

$$C_t^d = c_t^d \left[ \sum_{h \in [t, t+\Delta t]} (P_{n,h}^{dis} + P_{n,h}^{cha})\Delta h + q_t \right] \quad (4)$$

### B. Storage operation constraints

The energy constraints of storage are formulated as (5) and (6). The energy level of storage at time $h$, $E_h$, is a function of the energy level at time $h-1$, $E_{h-1}$, and the charging/discharging schedules at time $h$, as (5), where $\rho$ is the self-discharge rate, and $\eta$ is the charge/discharge efficiency. The energy level of storage cannot exceed its capacity, as (6), where $E_t^{max}$ is the energy capacity of the storage.

$$E_h = (1-\rho)E_{h-1} + \sum_{n \in \{n_A, n_B\}} \left( P_{n,h}^{cha}\eta_t\Delta h - P_{n,h}^{dis}\Delta h / \eta_t \right) \quad (5)$$
$$\forall h \in [t, t+\Delta t]$$

$$0 \leq E_h \leq E_t^{max} \quad \forall h \in [t, t+\Delta t] \quad (6)$$

The power outputs constraints are expressed as (7), where $P_t^{max}$ is the power capacity of the storage, and $\omega_{n,h}$ is a 0-1 variable that denotes whether the storage is at node $n$ and at time $h$ (1 indicates present and 0 indicates absent). We name $\omega_{n,h}$ as the location indicator. This indicator couples the operation and transportation constraints in this model.

$$0 \leq P_{n,h}^{dis}, P_{n,h}^{cha} \leq \omega_{n,h}P_t^{max} \quad \forall n \in \{n_A, n_B\}, h \in [t, t+\Delta t] \quad (7)$$

### C. Storage transportation constraints

The storage can only be present at one node at one time, which is formulated as (8). $\alpha_{n,h}$ is a 0-1 variable that denotes whether the storage is traveling to node $n$ at time $h$, while $\beta_{n,h}$ is a 0-1 variable that denotes whether the storage is traveling from node $n$ at time $h$.

$$\sum_{n \in \Omega_n} \omega_{n,h} \leq 1 \quad \forall h \in [t, t+\Delta t] \quad (8)$$

The travelling status of storage is modelled in (9) to (12), where $\alpha_{n,h}$ is a 0-1 variable that denotes whether the storage is traveling to node $n$ at time $h$; $\beta_{n,h}$ is a 0-1 variable that denotes whether the storage is traveling from node $n$ at time $h$; and $\theta_{n,h}$ is an 0-1 auxiliary variable.

$$\alpha_{n,h} - \beta_{n,h} = \omega_{n,h} - \omega_{n,h-1} \quad \forall n \in \{n_A, n_B\}, h \in [t, t+\Delta t] \quad (9)$$

$$\sum_{n \in \{n_A, n_B\}} (\alpha_{n,h} + \beta_{n,h}) \leq 1 \quad \forall h \in [t, t+\Delta t] \quad (10)$$

$$\sum_{n \in \{n_A, n_B\}} (\alpha_{n,h} - \theta_{n,h}) = \gamma_{h-1} - \gamma_h \quad \forall h \in [t, t+\Delta t] \quad (11)$$

$$\sum_{n \in \{n_A, n_B\}} (\alpha_{n,h} + \theta_{n,h}) \leq 1 \quad \forall h \in [t, t+\Delta t] \quad (12)$$

The travel time constraint is formulated as (13), where $T_h$ is the required driving and installation time for the storage to leave from one node and be prepared to operate at the other node, which may vary across time considering traffic congestion.

$$\gamma_h \geq \gamma_{h-1} - \gamma_{h-T_h} \quad \forall h \in [t, t+\Delta t] \quad (13)$$

## V. CASE STUDY

In this section, we present a case study for a truck of portable storage traveling between the pair of nodes around San Marcos mentioned in Section II. We optimize the operation and transportation strategies of a truck of portable storage rated at 2.7 MW/2.7 MWh doing spatiotemporal arbitrage between two nodes around San Marcos using model (1) to (13) for each day from October 2017 to September 2018. The time scale $\Delta h$ is set to 15 minutes. The storage is assumed to be a price-taker, which means that the outputs of the storage has not impact on the prices of the nodes. The real day-ahead energy prices are taken as the price forecasts. The charge/discharge efficiency is assumed to be 95%, which represents a round-trip efficiency of 90%. The self-discharge rate is assumed to be 0. For 5-mile single trip, given a $20 hourly wage for truck driver and a 25 miles/hour average speed, the travel time is 12 minutes, and the labor cost is $4.

The end of life of electrochemical energy storage is typically defined as when the remaining capacity decreases to 80% or 70% of the original capacity [20],[21]. In this paper, the storage utilization and degradation are measured by energy throughput (charging plus discharging) in megawatt hours in this paper. The cycle life of lithium-ion battery is assumed to be 3000 cycles, which is equivalent to 16.2 GWh. The calendar degradation rate for lithium-ion battery is assumed to be 1% capacity loss/year, which is equivalent to approximately 1.5 MWh energy throughput per day (assuming the storage life ends when the capacity has decreased to 70% of the initial). The optimal marginal cost of usage for portable storage is $25/MWh in this case, while $14/MWh for stationary storage, which are determined using an intertemporal operational decision framework [9]. The discount rate is assumed to be 7%.

Fig. 3 presents the optimal operation schedules of the portable storage and LMP profiles of the two nodes in a sample day. The broken lines represent the LMPs of the two nodes, respectively. From Hour 8 to Hour 21, there are remarkable price differences between the two nodes, which indicate congestions, with Node 1 being the high-price node and Node 2 being the low-price node. Although the real cause of congestion is unknown, it is reasonable to speculate that there is over-abundant solar generation at Node 2 according to the price profile. To exploit the price differences, the storage travels between the two nodes to discharge and sell energy at Node 1 (blue bars) and to charge and buy energy at Node 2 (orange bars). In Fig. 3, each time the bar color changes, the storage makes a trip from one node to the other, and we can see the storage makes 5 round-trips in the sample day. The travelling capability provides the portable storage much more profit opportunities compared to stationary storage, because the portable storage can profit from both price differences between different nodes and between different hours within one node, while the stationary storage can only profit from the price difference within one node. As seen from Fig. 3, the portable storage conducts 4 profitable cycles over the day, while for stationary storage it is only profitable to run 1-2 cycles during the peak and valley hours. The congestion between the two nodes can be relived as the storage brings over 9 MWh energy in total from Node 2 to Node 1 over the day.

TABLE II
MODEL PARAMETERS

| $P_t^{max}$ | $E_t^{max}$ | $\eta$ | $\rho$ | $c_{tr}$ | $T_h$ | $q_t$ |
|---|---|---|---|---|---|---|
| 2.7 MW | 2.7 MWh | 95% | 0 | $4/trip | 12 mins | 1.5 MWh |

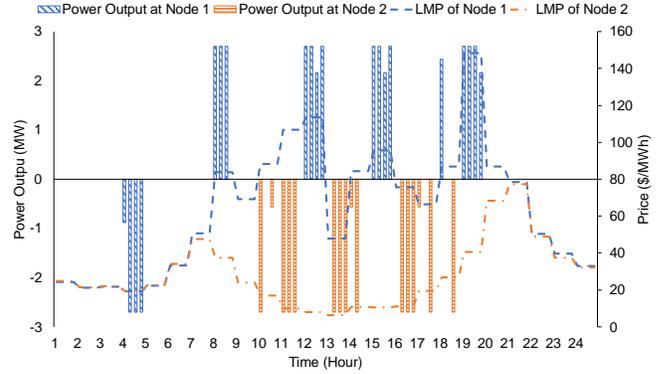

Figure. 3. Optimal operation schedules for a truck of portable storage doing spatiotemporal arbitrage between two nodes around San Marcos and the LMP profiles of the two nodes in a sample day (April 17, 2018).

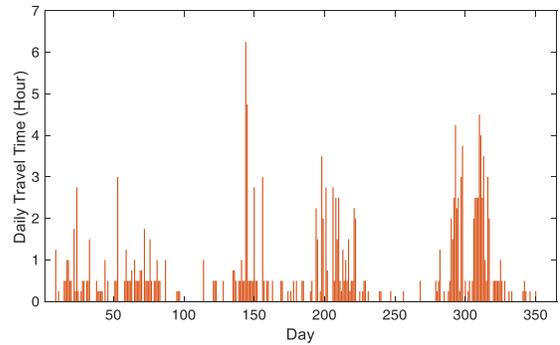

Figure. 4. Daily travel time for a truck of portable storage doing spatiotemporal arbitrage between two nodes around San Marcos from October 2017 to September 2018.

Fig. 4 presents the daily travel time for the portable storage over a year. The total travel time is 181 hours over one year,

TABLE III
COMPARISON BETWEEN PORTABLE STORAGE AND STATIONARY STORAGE

|  | Portable storage | Stationary storage |
|---|---|---|
| First-year revenue | $98,000 | $55,000 |
| Total life-cycle revenue | $548,000 | $308,000 |
| Extra life-cycle revenue of portable storage | $240,000 | — |
| Trucking cost | $150,000 | — |

which is approximately 0.5 hour per day. The results indicate that there are intensive travels during some periods of time, which coincides with the monthly price difference distribution in Fig. 2. In those days without significant price difference between these two nodes, the storage can serve other pairs of nodes with congestion in the neighborhood.

Table III compares the revenues between a truck of portable storage doing spatiotemporal arbitrage between the two nodes around San Marcos and a stationary storage of the same size doing arbitrage at Node 2. By trucking, the storage revenue in the first year is increased by $44,000, and the total life-cycle storage revenue is increased by $240,000, which is greater than $150,000, the cost of a Tesla Semi. This implies that for a stationary storage providing peak shaving or renewable integration, which is similar to energy arbitrage, it could be profitable for storage owner to make the storage portable through trucking.

## VI. CONCLUSIONS

This paper explores a new business model for energy storage, in which energy storage is loaded on truck and doing spatiotemporal arbitrage between nodes with congestion. We develop an optimization model for spatiotemporal arbitrage of portable storage and apply the model to simulate the operation and transportation of a truck of Tesla Semi loaded with Tesla Powerpack over two nodes in California. The results indicate that storage owner can earn net profits from converting stationary storage to portable storage in energy arbitrage application, and the transmission congestion can be relieved at the same time. The spatiotemporal arbitrage model can be extended to multiple applications (such as frequency regulation, voltage support, etc.), multiple nodes, and multiple trucks, which is one of our ongoing work.

One potential issue is that whether the storage can be assumed as price-taker or how to compensate the storage if it fully resolves the congestion and eliminates the price difference. Some technical issues such as safety and thermal control may also need attention in practice.

ACKNOWLEDGEMENT

This work was partially supported by the US Department of Energy under Grant DEEE0007165.